\documentclass[aps,showpacs,draft]{revtex4}

\usepackage{graphicx}

\newcommand{\Op}[1]{{\bf {\hat {#1}}}}
\begin{document}

\include{epsf}

\title{The Quantum Governor: Automatic quantum control and reduction of the influence of noise without measuring}
\author{S.\ Kallush, and R. Kosloff}
\affiliation{The Fritz Haber Research Center, The Hebrew University of Jerusalem,
Jerusalem 91904, Israel}

\begin{abstract}
The problem of automatically protecting a quantum system against noise in a closed circuit is analyzed. 
A general scheme is developed  built from two steps. At first, a distillation step is induced in which undesired components 
are removed to another degree of freedom of the system. Later a recovering step is employed which the system gains back its initial density. An Optimal-Control method is used to generate the distilling operator. The scheme is demonstrated by a simulation of a two level byte influenced by white noise. Undesired deviations from the target were shown to be reduced by at least two orders of magnitude on average. 
The relations between the quantum version of the classical Watt's Governor and the field of quantum information are also discussed.
\end{abstract}

\pacs{PACS Numbers: }

\maketitle

\section{Introduction}
Watt's Governor (WG), which was built in 1782, might be the very first machinery to solve automatically a control 
problem. As an automatic control tool, Watt's governor aims to conserve some physical properties 
of a system subject to stochastic noise while maintaining its internal dynamics.
Schematically, the WG can be viewed as a two step process, measurement and correction. 
At the first step a measurement of the system is performed to check whether 
the constraint has been violated. Next, if such a violation was found, a correction step takes place,
and drives the system back to the allowed boundaries.

A quantum Governor is a natural requirement when the limit of nano machines is approached.
Quantum computing \cite{Steane,DiVincenzo1,Lloyd,Nielsen} is another candidate for such a device.
However, quantum mechanics imposes non trivial restrictions on the development of the quantum Governor. A measurement, which is a main feature of the classical WG  intervenes in the dynamics 
of quantum systems and it therefore  should be avoided or reduced to a minimum.

A control scheme is traditionally categorized either as an open-loop or a closed-loop control
\cite{ZhouDoyGlo,Brogan}.
In the closed-loop control scheme one tries to extract information from a feedback from the quantum system 
in a way that allows the control of the system\cite{Doherty,Mabupop}. 
Within this scheme, a controlled collapse of some of the wavefunction occurs and converts some of the 
quantum variables of the system into classical parameters. One needs therefore to delicately 
balance the amount of withdrawn information in order to conserve the quantum character of the system.
Feedback control of quantum systems has being extensively studied during recent years by several groups
\cite{Doherty,Mabupop,Berglund,Steck}. 

An open-loop control scheme corrects the system without any measurement. In order to 
realize the  QG a tool that distills quantum systems in some automatic fashion, has to be built. 
Distillation steps usually reduce the density of the system, and hence, in order to conserve 
the density of a controlled system, an extra-step to enrich the system and compensate for 
the losses is also required.

In this paper we suggest and demonstrate a physical realization for an open-loop QG scheme.
Our open-loop QG is a two step routine (Cf. Fig. 1 ).
The scheme starts from an  initial state, for example a diatomic 
molecule in its ground electronic and vibrational state. This state is then 
disturbed by noise. At the first stage of action, an external field is applied to the 
system with the purpose of distilling the undesired components. The rejected components are moved 
to another degree of freedom of the system. In the present model  to an excited electronic state. 
At the next stage, the freely evolving dynamics is set
to enrich the initial state and return it back to the initial density. 

\begin{figure}[tbp]  
 \centerline{\epsfxsize=4.25in\epsfclipon\epsfbox{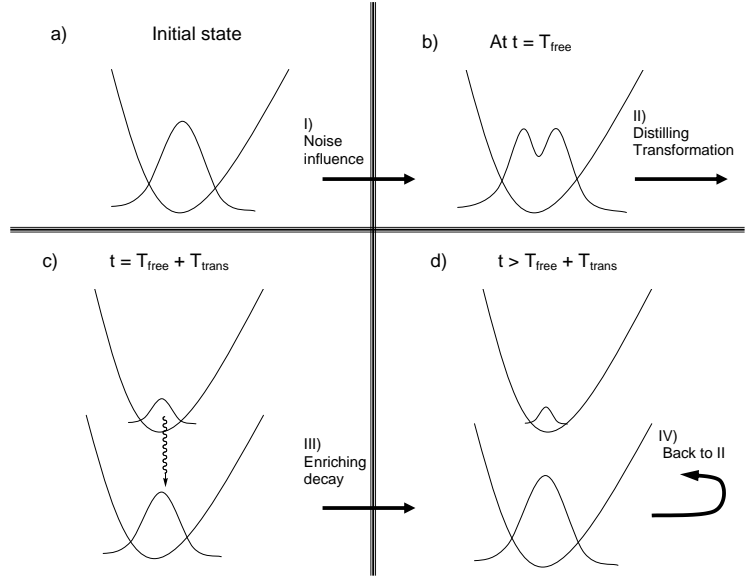}}
 \caption{Schematic principles of the QG: (a) An initial state is propagated under the influence of noise. (b) The initial state after the distortion. (c) The undesired part of the state is removed to another degree of freedom by a controlled transformation, here to another electronic surface. The initial state is purified but its density is reduced. (d) A properly designed decay transfers population back to the initial state and reconstruct its density during a free evolution of the system.}
 \label{fig1}
\end{figure}

We will show that the distillation step can be performed by a particular unitary transformation.
Recently, Optimal-Control-Theory (OCT) has been applied to find
the field generating the unitary operation \cite{Joseprl,Josepra,Sklarz}. 
In the present paper the methods of refs. \cite{Joseprl,Josepra} 
are generalized to obtain the control field for non-unitary
operations under non-unitary time-evolution governed by the Liouville equation for open quantum systems.
This control field is then employed to derive the distillation stage of the QG. This step
is followed by a field-free propagation of the system leading to the desired reconstructed state.

In this paper the QG model is presented and the tools required for its construction. 
In Section (II) a simple QG model is presented. Section (III) elaborates on the OCT mathematical considerations required to achieve the distilling transformation. Section (IV) presents  simulations of the QG. A discussion and conclusion is presented in section (V).

\section{The QG Model}
The Hamiltonian of the quantum Governor  QG model is partitioned to:
\begin{equation}
{\Op H} = {\Op H^0} + {\Op H_{noise}} + {\Op H_{G}}
\end{equation}
where, ${\Op H^0}$ is the free Hamiltonian of the system, ${\Op H_{noise}}$ is the stochastic noise, and ${\Op H_{G}}$
is the control part, governed by an external field.
The automatic control scheme could be applied  to complex quantum
systems. The principles of such a control 
scheme will be demonstrated by a simplified model
composed of a single two-level qubit with frequency $\omega_{1}$:
\begin{equation}
{\Op H^0}  = \left[ {\begin{array}{*{20}c}
   0 & 0  \\
   0 & {\omega _1 }  \\
\end{array}} \right]
\end{equation}
with the two states denoted by $\left\{ {\left| 1 \right\rangle _g ,\left| 2 \right\rangle _g } \right\}$.
The two levels can represent for example two 
spin states or two vibrational levels in a diatomic molecule. 
The qubit is then influenced by an external noisy field:
\begin{equation}  
{\Op H}_{noise}  = \mu f\left( t \right) \left[ {\begin{array}{*{20}c}
   0 & 1   \\
   1  & 0  \\
\end{array}} \right]
\label{noise}
\end{equation}
where $\mu$ is the dipole moment and $f(t)$ is a white noise function which obeys:
\begin{equation}
\left\langle {f\left( t \right)} \right\rangle  = 0\,\,\,\,\,\,\,\left\langle {f\left( t \right)f\left( {t'} \right)} 
\right\rangle  = \zeta \delta\left( t-t' \right)
\end{equation}
The target of the control is to conserve an initial qubit state protecting it from the noise. We first will describe the route to build a QG for a particular target byte. This approach will then be extended to  a general target byte.

\subsection{The Conservation of a byte in its ground state}

The state of the system is described by a density operator in the energy representation. The  target 
and initial state are chosen as: 
\begin{equation}
{\Op \rho ^0 } = \left| 1 \right\rangle \left\langle 1 \right| = \left( {\begin{array}{*{20}c}
   1 & 0  \\
   0 & 0  \\
\end{array}} \right)
\label{groundinitial}
\end{equation}
The propagation in time of ${\Op \rho ^0}$ under the influence of the noise leads to an undesired population on the excited $\left| 2 \right\rangle $ state. To restore the state, the qubit is coupled to an auxiliary
qubit with frequency $\omega_{2}$, $\left\{ {\left| 1 \right\rangle _e ,\left| 2 \right\rangle _e } 
\right\}$.
The second qubit can be realized for example as  two vibrational levels within the excited electronic state.  The distillation step is achieved by applying the unitary swap transformation:
\begin{equation}
{\Op O_{d} } = \left( {\begin{array}{*{20}c}
   1 & 0 & 0 & 0  \\
   0 & 0 & 1 & 0  \\
   0 & 1 & 0 & 0  \\
   0 & 0 & 0 & 1  \\
\end{array}} \right)
\label{targetO}
\end{equation}
The outcome of this transformation is that all the undesired population is transferred to 
the $\left| 1 \right\rangle_e$ of the auxiliary $e$ byte. 
This step cancels also the phase between the two states (see Fig. 2a).

\begin{figure}[tbp]  
 \centerline{\epsfxsize=3.7in\epsfclipon\epsfbox{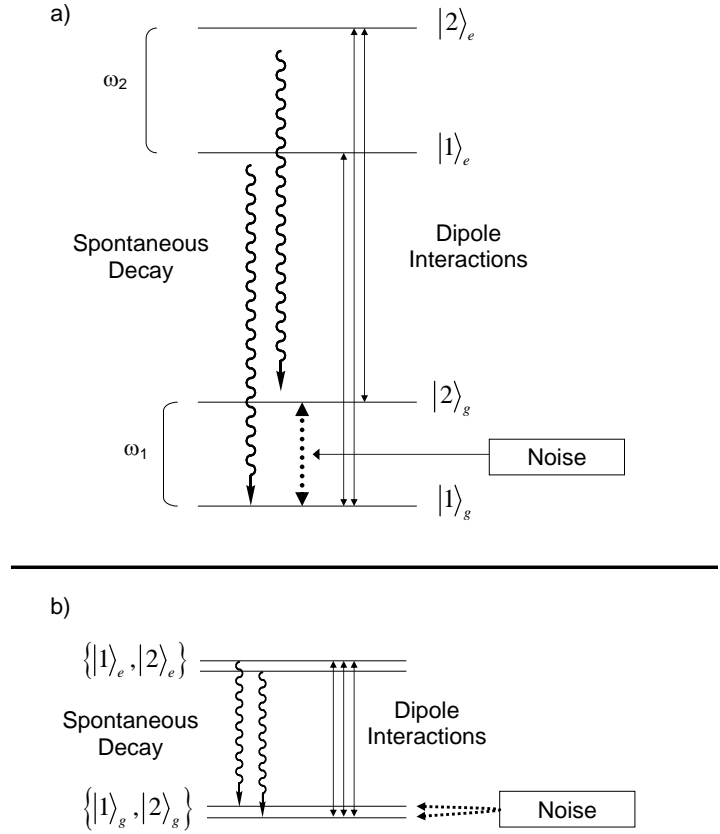}}
 \caption{Level scheme for a QG. (a) QG which conserves a byte on its ground state. A two level system with frequency $\omega_1$ is influenced by noise. An external field can be applied on the byte to couple it with another byte with frequency $\omega_2$. Spontaneous decay is allowed only as depicted on the figure. (b) QG which conserves a general byte. The lower (target) byte is composed of two degenerate levels, and so is the upper (control) byte. The external field coupled between the two bytes, while the spontaneous decay is allowed only as depicted.}
 \label{fig2}
\end{figure}
  

Note that the transformation is done under Liouville evolution which allow also non-unitary transformations to take place. A possible one step solution for the QG might be to leave the population in the desired state untouched while moving all the other population to the desired state, e.g, with the distiling operator
\begin{equation}
{\Op O} \propto \left( {\begin{array}{*{20}c}
   1  & 0  \\
   1  & 0  \\
\end{array}} \right)~~.
\end{equation}
This will use heavily the uncontrolled, non-unitary components of the Liouvillian operators. Working at the weak dumping limit and demanding a fast transformation
turn this idea to be unpractical.
We are therefore forced to use a transformation
which is close to unitary, and add another step for the completion of the task.

The free evolution step is generated by the following Liouville-von Neumann equation:
\begin{equation}
\frac{{\partial {\Op \rho} }}{{\partial t}} =  - \frac{i}{\hbar }\left[ {{\Op H}^{0}+{\Op H}_{noise},{\Op \rho} } \right] + {\cal{L}}_D \left( {{\Op \rho} } \right)
\label{liou}
\end{equation}
Where ${\cal{L}}_D$ is a particular dissipative  Liouville superoperator. 
 
${\cal L}_D$ induces a selective decay between the vibrational states: 
$\left| 1 \right\rangle_e \to \left| 1 \right\rangle _g $ and $\left| 2 \right\rangle _e \to \left| 2 \right\rangle_g $. 
For example a decay induced by the transition state dipole between vibronic (vibrational+electronic) 
which has the above selection rules.

After the distillation step, both the lower and upper bytes reside on the target state within the single byte subspace, i.e., with the correct relative population and phase. The decay step then restores the
population from the upper byte reconstructing the density of the target byte. A cyclic application of this scheme on a quantum system purifies it reducing the  influence of the noise to  minimum. 

This process can also be considered as a perpetual preparation of the desired state. Note, however, that this desired state itself is never achieved directly in any of the steps.  The governor acts by enforcing the conservation of the state  by properly aligning the state on the directions of the target.   

\subsection{The Conservation of a byte in a general state}

We next study the preservation of a superposition state which has the form:
\begin{equation}
\left| { +  } \right\rangle  = a\left| 1 \right\rangle _g  + b\left| 2 \right\rangle _g 
\end{equation}
with $|a|^2 + |b|^2 = 1$.

The governor utilizes the unitary transformation:
\begin{equation}
{\Op U}= \left( {\begin{array}{*{20}c}
   a & {b^* } & 0 & 0  \\
   b & { - a^* } & 0 & 0  \\
   0 & 0 & a & {b^* }  \\
   0 & 0 & b & { - a^* }  \\
\end{array}} \right)
\end{equation}
This unitary operator transforms the basis set from the original basis
$\left\{ {\left| 1 \right\rangle _g ,\left| 2 \right\rangle _g ,\left| 1 \right\rangle _e ,\left| 2 \right\rangle _e } \right\}$
to a new basis set $\left\{ {\left|  +  \right\rangle _g ,\left|  -  \right\rangle _g ,\left|  +  \right\rangle _e ,\left|  -  \right\rangle _e } \right\}$.
Now, after the noise influences the initial byte, the original byte can be distilled by the transformed 
form of the swap operator ${\Op O_{d}}$, Cf. Eq. (\ref{targetO}):
\begin{equation}
{\Op O}_{d}^{g}  = {\Op U}  {\Op O}_{d} {\Op U }^\dag= \left( {\begin{array}{*{20}c}
   {\left| a \right|^2 } & {ab} & {a^* b} & {b^2 }  \\
   {a^* b^* } & {\left| b \right|^2 } & { - a^{*2} } & { - a^* b}  \\
   {ab^* } & { - a^2 } & {\left| b \right|^2 } & { - ab}  \\
   {b^{*2} } & { - ab^* } & { - a^* b^* } & {\left| a \right|^2 }  \\
\end{array}} \right)
\label{eq:uniti2}
\end{equation}
So that under the operator ${\Op O}_{d}^{g}$:
\begin{equation}
\begin{array}{l}
 {\Op O}_{d}^{g} \left|  +  \right\rangle _g  = \left|  +  \right\rangle _g  \\ 
 {\Op O}_{d}^{g} \left|  -  \right\rangle _g  = \left|  +  \right\rangle _e  \\ 
 {\Op O}_{d}^{g} \left|  +  \right\rangle _e  = \left|  -  \right\rangle _g  \\ 
 {\Op O}_{d}^{g} \left|  -  \right\rangle _e  = \left|  -  \right\rangle _e  \\ 
 \end{array}
\end{equation}
It has been noticed \cite{Bacon,Dyakonov} that,
non-degenerate qubits are very difficult to handle due to the relative
coherent phase that develops under the free evolution. To avoid this problem 
we take the two bytes for the conservation of a general byte as a two couples of degenerate states. 
Initially all of the population is on the target byte. After the distilling transformation 
both the target and the auxiliary bytes are, within the single byte subspace, on the desired state, i.e.: 
$\left|  +  \right\rangle _{g/e} $, with the right phase between the two states of each of the bytes.

The transformation of Eq. (\ref{eq:uniti2})  corrects the error in both population and phase. The principle of the correction is to move a relative error between two states {\it within} a single byte, 
to a relative error between two bytes. A decay step must then return back the population 
from the control byte to the target byte and annihilate the relative errors between the bytes.
This task is achieved using the same assumption used previously with the allowed transitions $\left| 1 \right\rangle_e \to \left| 1 \right\rangle _g $ and $\left| 2 \right\rangle _e \to \left| 2 \right\rangle_g $, while other possibilities are forbidden. 
The scheme for the general byte conservation is illustrated at the lower panel of Fig. 2.

\section{Optimal Control Theory for Non-Unitary Transformation 
Under Non-Hamiltonian Dynamics}

The quantum governor is achieved by the unitary transformations responsible
for the distillation.  The next step is to find the external field that induces such a transformation.
This task is achieved by an inversion process which starts from the unitary operator responsible for
the distilling and determines the field. The present description follows the treatment of Ref. \cite{Bartana}. 
A target transformation for a N-level space is described by 
the N-by-N matrix representing  the operator ${\Op O}$. ${\Op O}$ is neither necessarily unitary, 
or orthogonal. Nevertheless, practically it cannot deviate too much from unitarity.
Our target is to find the field that generates the transformation ${\Op O}$ at time $t = T_{trans}$, 
independent of the initial state. 

The  density operator is now decomposed into  a sum of a complete 
basis set of operators in the Hilbert-Schmidt space. 
The complete set for a N-level system density operators contains $2^N$ hermitian matrices of dimension N-by-N. 
A scalar product between two operators ${\Op A}$ and ${\Op B}$ in Hilbert-Schmidt space is defined as\cite{Nielsen}:
\begin{equation}
\left( {\Op A}  \cdot {\Op B} \right)   = Tr\left\{ {\Op A}^\dag  {\Op B}  \right\}
\end{equation}
The norm of an operator is therefore: $|{\Op A}|  = Tr\left\{ {{\Op A} ^\dag  {\Op A} } \right\} $.
For density operators: $ \frac{1}{N} \le |{\Op \rho}|  \le 1 $, so that $ |{\Op \rho}| =1 $ for a pure state, and $ |{\Op \rho}|= 1/N $ 
for the classical limit for $N$ degenerate states.
Note that under unitary dynamics the norm and the entropy of a density operator are conserved.
This, however, is not true under dissipative conditions.

For a complete base of operators $\left\{{\Op G}_j^0 \right\}$
the desired operation ${\Op O}$, changes each of the basis set operators to new target operator:
\begin{equation}
{\Op G}_j^{\rm{target}}  = {\Op O} {\Op G}_j^0 {\Op O}^\dag
\end{equation}
Under unitary transformation the complete orthonormal set is transformed to another complete orthonormal set. 
This is not true for non-unitary transformations. 
A chosen functional for the optimization procedure should reflect deviations between the propagated operators and 
the target set of operators. The set of operators $\left\{{\Op G}_j^{\rm{target}}\right\}$ does not conserve the initial norm. 
Therefore is better to define the functional by:
\begin{equation}
\tilde{F} = \sum\limits_{j} \left( {{\Op G}_j^0  \cdot {\Op G}_j^{result} } \right)
\end{equation}
where the set $\left\{ {\Op G}_j^{\rm{result}} \right\}$ is obtained by propagating the 
set of target operators $\{ {\Op G}_j^{\rm{target}} \}$ {\it backward} in time. When the target is achieved $F = 2^N $.

Two additional constraints are imposed:
\begin{enumerate}
\item { The reverse time-evolution of the system is also governed by the
Liouville-von Neumann equation.}
\item{The total field energy has to be minimized.}
\end{enumerate}
To meet these demands a modified functional is employed:
\begin{equation}
\begin{array}{l}
F = \tilde{F} - \sum\limits_j {\int\limits_{T_{trans} }^0 \left( {\left[ \frac{\partial {\Op G}_j }{\partial t} - {\cal L}^{*}  \left( {\Op G}_j  \right)\right]  \cdot {\Op B}_j }\right) dt}  \\ 
  - \int\limits_{T_{trans}}^0 {\lambda \left( t \right)\left| \varepsilon  \right|^2 dt}   \\
\end{array}
\end{equation}
${\Op B}_j$ are $2^N$ operator Lagrange multiplier, and $\lambda(t)$ is a time-dependent scalar Lagrange multiplier.
An extremum for $F$ is obtained by a variation of $F$ with respect to $\delta {\Op G}_j$ and the field.
Following Ref. \cite{Bartana}, the equations of motion for the
reverse propagation of the $G_j$'s become:
\begin{equation}
\frac{ \partial {\Op G}_j }{\partial t} =  + \frac{i}{\hbar}\left[ { {\Op H},{\Op G}_j } \right] + {\cal{L}}_{D}^{*}  \left( {\Op G}_j  \right)
\end{equation}
with the initial conditions ${\Op G}_j(T_{trans}) = {\Op G}_j^{\rm{target}}$, and for the forward time-propagation of the ${\Op B}_j$ operators:
\begin{equation}
\frac{{\partial {\Op B}_j }}{{\partial t}} =  - \frac{i}{\hbar}\left[ { {\Op H},{\Op B}_j } \right] + {\cal{L}}_D \left( {\Op B}_j  \right)
\end{equation}
and $B_j(0) = G_j^0$. 
Application of the Krotov's iterative method to obtain monotonic increase toward the objective at each iteration defines 
the field at each new iteration at time $t$:
\begin{equation}
\epsilon^{new}(t)  = \epsilon^{pre}(t)  + \lambda C(t) \sum\limits_j {\left( {\left[ {\Op \mu} ,{\Op G}_j(t) \right]} \cdot {{\Op B}_j(t)}   \right)} 
\end{equation}
$C(t)$ is a time dependent envelope function, usually a Gaussian, 
and $\lambda$ is a strategy parameter. 
Large values of $\lambda$ will cause rapid changes of the field at each iteration. 
The difficulty in building quantum computing components is believed to grow exponentially with the number 
of bytes for systems which evolve under unitary dynamics. 
For non-unitary dynamics it is expected to be even worse. 
This expected result originate  from the fact that under non-unitary dynamics 
there are more options for decoherence to occur. 

\section{Simulation and Results}
\subsection{Protecting a byte on its ground state}

The application of the QG in protecting a target byte of the form of Eq. (\ref{groundinitial}) is now demonstrated. 
The first step of the procedure is to calculate the field required for the generation of the distilling transformation ${\Op O}_d$ Cf. Eq. (\ref{targetO}).
Two main timescales dominate the QG model: $1)\, \tau_{trans}$ - the time duration for the field-derived transformation,
and $2)\, \tau_{free}$ - the period of free propagation, where $\tau_{trans} \ll \tau_{free}$. 
The characterizing parameters of the four levels of the model are taken as the two couples of the 
two lowest vibrational states within the first two electronic states of the 
Na$_2$ molecule. Table \ref{param} summarizes the parameters used in the simulations.

\begin{table}
\begin{tabular}{|c|c|}
	\hline
   Parameter    &   Value     \\
	\hline
$\Delta $ & $0.06601$ Hartree \\
$\hbar\omega_1$ & $7.2449716268\times 10^{-4}$ Hartree \\
$\hbar\omega_2$ & $5.3746313155\times 10^{-4}$ Hartree \\
$\tau_{trans}$ & $1.08$ps\\
$\Gamma^{-1}$ & $10.0$ps \\
$\tau_{free}$ & $241$ps \\
	\hline
	\end{tabular}
\caption{Parameters used in the simulation: $\Delta$ is the electronic energy gap, $\omega_j$ 
is the vibrational frequency for the $j-$th
electronic state, $\tau_{trans}$ is the duration of the distillation transformation, 
$\Gamma$ is the decay rate, and $\tau_{free}$ is the period of a correction cycle.}
\label{param}
\end{table}

The operation are carried out in a 4-level space. Due to the decay of the upper byte, it can be assumed that when the 
controlled transformation is applied, all the population is at the lower byte. 
Hence, only a basis set for a single byte  ,i.e., the lower byte, operations subspace within a four level system is needed. 
The time evolution generated by the Liouville Van Neumann equation was calculated
by the Newton polynomial expansion method\cite{Huisinga}.

Fig.~3 displays the infidelity of the transformation, $ \log(1 - F/2^{n})$ vs. the number of iterations. $n$ 
is the number of basis states involve in the operation, here $2$. 
The two inset panels of fig.~3 show the resulting fields and their Fourier transforms which generate
an infidelity close to -12. 
The two peaks in the frequency domain are the result of the initial guess which initiated the 
optimization process: $\epsilon(t) \propto \cos(\Delta t)$ where $\Delta$ is the vertical energy gap between the two electronic states. 

\begin{figure}[tbp]  
 \centerline{\epsfxsize=3.7in\epsfclipon\epsfbox{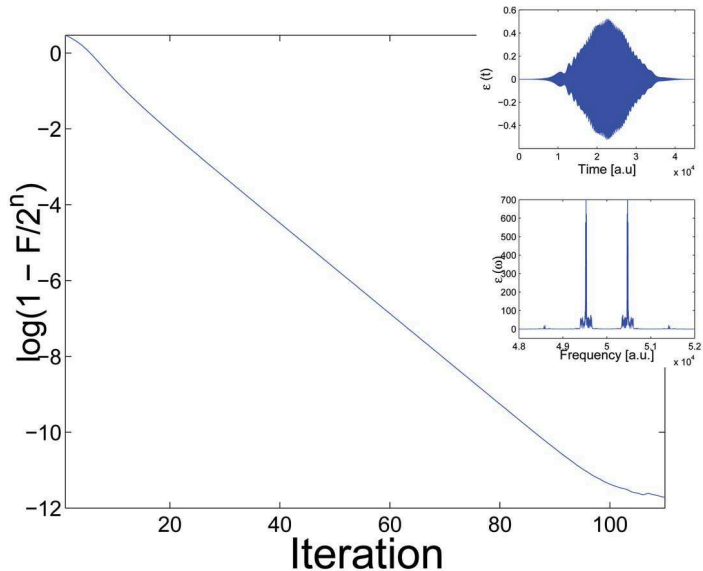}}
 \caption{Main frame: Infidelity of the transformation gate ${\Op O} _{d}$ versus the number of the iterations. 
 Upper and lower inset panels: time and frequency dependency of the field.}
 \label{fig3}
\end{figure}
At the second step, the target byte $\rho^0 = \left| 1 \right\rangle _g \left\langle 1 \right|_g $
is propagated freely for a duration $\tau_{free}$ with the additional distilling transformation. 
The white noise function for the propagation Cf. Eq. (\ref{noise}) is modeled as:
\begin{equation}
f\left( t \right) = N \, s\left( t \right)
\end{equation}
where N is the noise intensity and $-0.5 < s\left( t \right) < 0.5 $ is generated from a uniform random distribution. 
The decay rates are taken to be $\Gamma_{11}^{-1} = \Gamma_{22}^{-1}$. For comparison, three calculations 
were performed with the same noise parameters:
\begin{itemize}
\item{A reference propagation: the system is propagated with noise without any correction.
The red line in Fig.~4 represents the value of $R=1 - \left({\Op \rho} ^0  \cdot {\Op \rho} (t)\right)$ versus time. 
A significant deviations from the initial  byte develops, up to $R=0.02\%$,  due to the noise. 
The time-averaged deviation is about one third of the maximal deviation.}
 \item{A partially controlled propagation: the distilling transformation is applied on the byte each $\tau_{free}$, but no decay 
 between the bytes was allowed. The black line in Fig. 4 displays the deviations of the density matrix from the target ${\Op \rho}^0$ vs. time, for this case. According to the transformation ${\Op O}_d$ any remains of population on the upper byte will be transferred back to the lower byte and ruin the efficiency of the noise reduction. As expected, the growth of deviations from the target byte develops here almost at the same rate as in the previous case. The maximal deviation is $0.01\%$ and the average is again one third of the maximal value.}
\item{Fully Controlled propagation: the system is propagated with both the distilling transformation and the 
decay period between the bytes. The blue line in Fig. 4 represents  deviations of the density matrix from the target
for the fully controlled propagation. It can clearly be seen that the full scheme works well. 
The maximal deviation under the fully controlled propagation is reduced to $7 \times 10^{-4}\%$ 
and the average to approximately  $6 \times 10^{-5}\%$. Figure 5 is a blowup of this line. 
The influence of the very frequent corrections is  clearly visible. 
The system is more stable by two orders of magnitude on average under the QG scheme.}
  \end{itemize}
  \begin{figure}[tbp]  
 \centerline{\epsfxsize=3.75in\epsfclipon\epsfbox{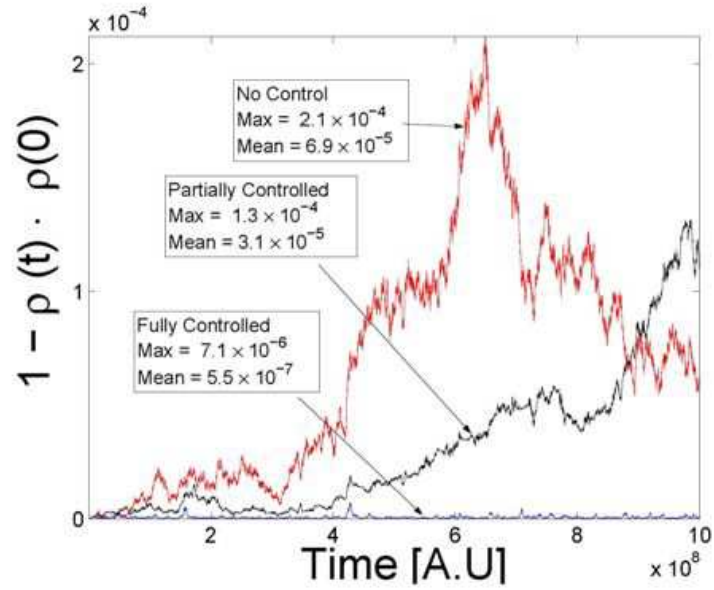}}
 \caption{The deviation of the byte vs. time for the uncontrolled (red), partially controlled (black), 
 and fully controlled (blue) cases. The deviation is defined as $1 - \left({\Op \rho} ^0  \cdot {\Op \rho} (t)\right)$. 
 See the text for the parameters used for the simulation.}
 \label{fig4}
\end{figure}
  \begin{figure}[tbp]  
 \centerline{\epsfxsize=4.25in\epsfclipon\epsfbox{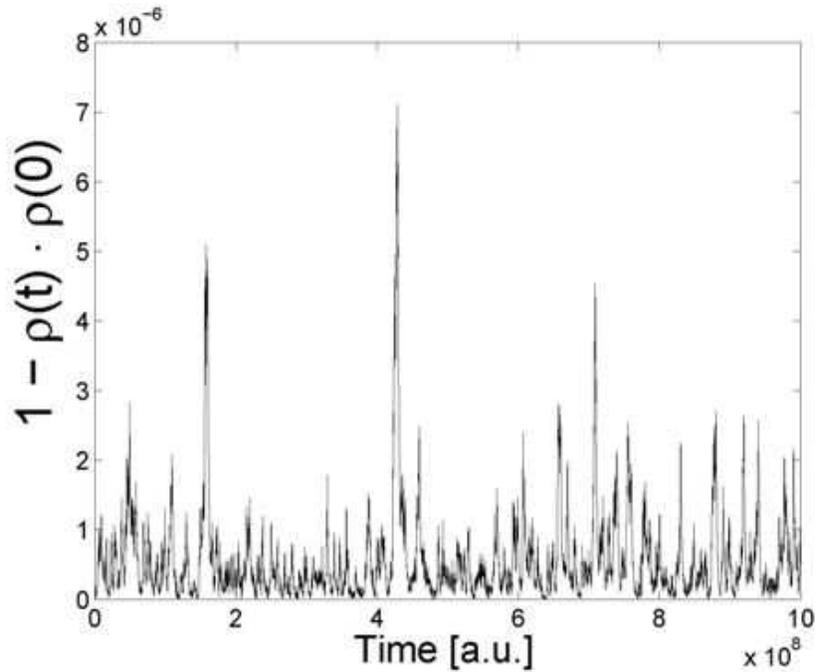}}
 \caption{A blowup of the deviation of the byte vs. time for the fully controlled system: A distilling transformation is apply every $\tau_{free} = 241$ps.}
 \label{fig5}
\end{figure}

Changes of some of the parameters of the system, e.g., taking $\tau_{free}$ to be smaller, can stabilize the system even further.

\subsection{Conservation of a general byte and an investigation of the necessary features for a QG}

For a general byte case the normalized state was chosen as the target byte:
\begin{equation}
\left|  +  \right\rangle _g  = \left|  +  \right\rangle  = \frac{1}{{\sqrt {1 + b_n^2 } }}\left( {\left| 0 \right\rangle  - ib_n \left| 1 \right\rangle } \right)
\end{equation}
and $b_n$ randomly chosen to be equal to $0.23113851357429$. The two bytes were chosen as two pairs of degenerate TLS. 
The optimal field for this transformation gate was converged to the same accuracy as at the previous section
with approximately the same speed of convergence\footnotetext[1]{Several numerical tests were performed for the present case as well as
for the case of unitary transformation under unitary time evolution of ref. \cite{Josepra}. Typically the difficulty to achieve the optimal field depended very weakly on the nature of the desired gate.}. 

The noise model influenced directly both the population and the phase between the states:
\begin{equation}
{\Op H}_{noise}  = \mu f\left( t \right)\left[ {\begin{array}{*{20}c}
   -1 & 1   \\
   1  & 1  \\
\end{array}} \right]
\end{equation}

To gain more insight on the necessary features needed for a QG to work properly  several simulation for mutated QGs are presented. 
Two conventions for the characterizing parameters are employed: the first one is identical to the previous one:
\begin{equation}
 R = 1 - \left( {\Op \rho} ^0  \cdot {\Op \rho} ( t) \right).
 \end{equation}
A scheme which conserves low values of $R$ can be defined as a fully conserved scheme. 
The distilling transformation aligns the byte to the right direction, i.e., with the correct relative phase and population between the two states. A byte can deviated from the target byte in its norm but still conserve a high resemblance to it with respect to the correct phase and population. A normalized deviation is defined as:
\begin{equation}
R_n  = 1 - \frac{\left({{\Op \rho} ^0  \cdot {\Op \rho} \left( t \right)}\right)}{{\left| {{\Op \rho} \left( t \right)} \right|}} = 1 - \frac{\left({{\Op \rho} ^0  \cdot {\Op \rho} \left( t \right)}\right)}{\sqrt{ \left({ {\Op \rho} \left( t \right) \cdot {\Op \rho} \left( t \right)} \right)}}~~~.
\end{equation}
The upper and the lower panels of figure 6 represents the measures $R$ and $R_n$ for five simulations with the same noise
parameters. The results are concentrated in table\ref{results1}.
\begin{figure}[tbp]  
\centerline{\epsfxsize=3.8in\epsfclipon\epsfbox{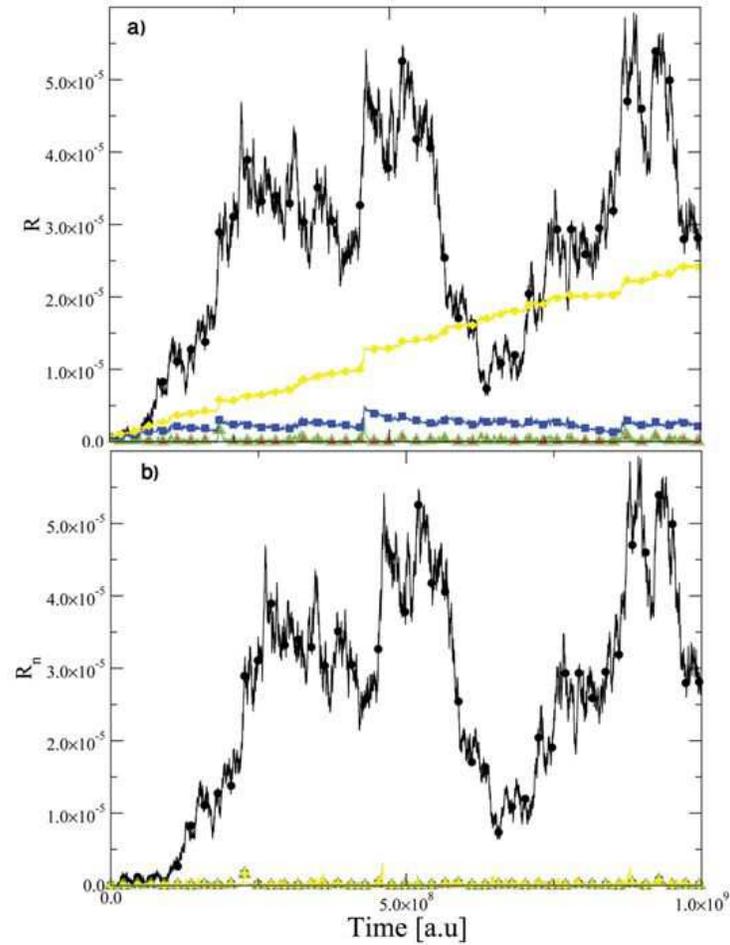}}
\caption{The deviation of a general byte vs. time for the cases of uncontrolled propagation (black), equal (red), different (green) and exchanged (blue) decay rates, and for the case where the decay channel is not directed back to the subspace of the controlled byte(yellow). The two panels display (a) $R$ - the un-normalized and (b) $R_n$ - the normalized deviations, respectively. See in the text for a more detailed description of the various cases}
 \label{fig6}
\end{figure}

 \begin{figure}[tbp]  
 \centerline{\epsfxsize=3.8in\epsfclipon\epsfbox{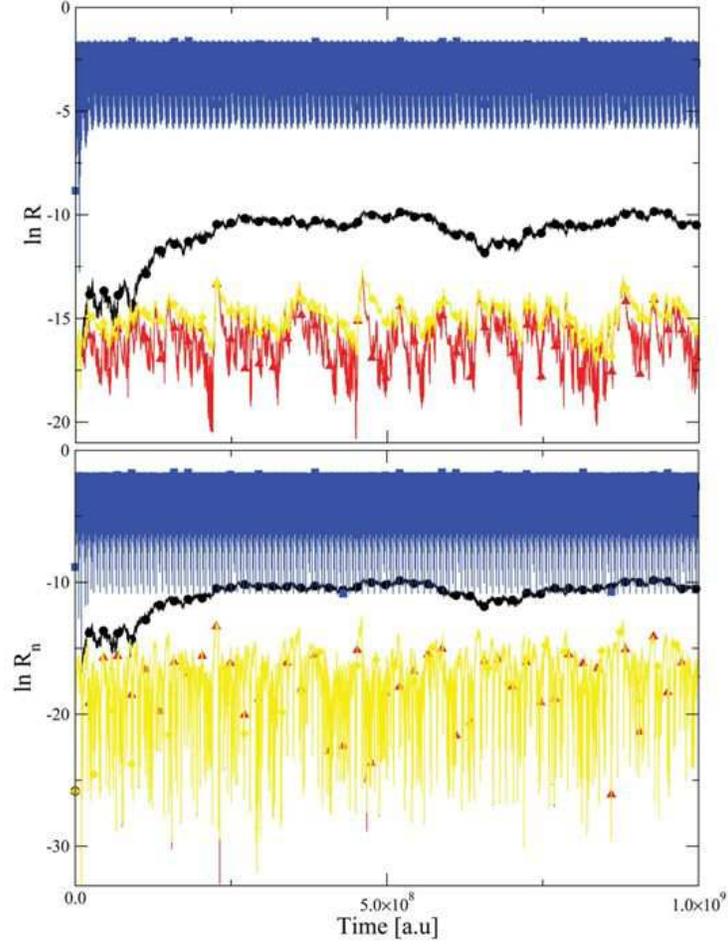}}
\caption{Logarithm of the deviation of a general byte vs. time for the cases of uncontrolled propagation (black), equal decay rates (red), non-degenerate byte (green) and scrambled transformation (yellow). 
The two panels display (a) $R$ - the un-normalized and (b) $R_n$ - the normalized deviation. See in the text more detailed description of the various cases}
 \label{fig7}
\end{figure} 

 \begin{table}
\begin{tabular}{|c|c|c|c|c|}
	\hline
 &  $R^{max}$    &   $R_n^{max}$  & $
\left\langle {R } \right\rangle $   & $
\left\langle {R_n } \right\rangle $
  \\
	\hline
1   & $5.9\times 10^{-5}$ & $5.9\times 10^{-5}$ & $2.9\times 10^{-5}$  & $5.9\times 10^{-5}$ \\
2   & $3.1\times 10^{-6}$ & $3.1\times 10^{-6}$ & $2.0\times 10^{-7}$  & $1.3\times 10^{-7}$ \\
3   & $3.1\times 10^{-6}$ & $3.1\times 10^{-6}$ & $2.0\times 10^{-7}$  & $1.3\times 10^{-7}$ \\
4   & $4.8\times 10^{-6}$ & $3.1\times 10^{-6}$ & $2.2\times 10^{-6}$  & $1.3\times 10^{-7}$ \\
5   & $2.4\times 10^{-5}$ & $3.1\times 10^{-6}$ & $1.2\times 10^{-5}$  & $1.3\times 10^{-7}$ \\
	\hline
	\end{tabular}
	\caption{Deviations from a general target byte for five different test cases (see the text for details). The columns display the maximal and the average of the deviations $R$ and $R_n$ for the five cases. }
	\label{results1}
\end{table}

Five test cases are simulated: 
\begin{enumerate}
\item{A reference propagation without any correction depicted by a black line with  closed circles. $R =R_n$}
\item{A fully controlled scheme with equal decay rate. $\Gamma_{11}^{-1} = \Gamma_{22}^{-1} = 10ps$, is depicted 
by the red line with the closed triangles.  The efficiency of the QG is the same as for previous case of the ground state byte. 
Both $R$ and $R_n$ are conserved to values well below the un-corrected propagation. The system is more stable by a factor of $20$ at the worst case and by more then a factor of $200$ on average.}
\item{This scheme is identical to (2), but with different decay rates, $\Gamma_{11}^{-1} = \frac{1}{10\pi}\Gamma_{22}^{-1} = 10ps$. 
Only selective decay is allowed. The results are depicted by the green line with open triangles. 
The scheme seems to work at the same efficiency as the previous case. 
The relative phase that develops {\it between} the two bytes during the propagation due 
to their different energies is destroyed by the decay  so that only the inner phase between the states is conserved. }
\item{In this case the decay channels were switched so that the decay channels are $\left| 1 \right\rangle_e \to \left| 2 \right\rangle _g $ and $\left| 2 \right\rangle _e \to \left| 1 \right\rangle_g $. The results (blue line with closed squares), show that under this scheme the noise is accumulated and the un-normalized deviation from the target is constantly growing. However, the accumulated error is still bearable, so that the mutated QG manages to transform the undesired part to the other byte and $R_n$ is still conserved.}
\item{In this test case (yellow line with closed diamonds), it was assumed that 
the population of the upper byte decays by some drain channel to a bath outside of the system. 
Under this scheme the remaining byte is well conserved   
but the norm of the state is reduced significantly.}
\end{enumerate}

Another four test-cases simulations are presented in figure 7. The numerical results are summarized in table\ref{results2}. 
The lines with black circles and red triangles present as on the previous demonstration, the uncontrolled 
and fully controlled QG. The case presented by the lines with the blue squares has a non-degenerate upper byte. 
The non degenerate states of the byte develops a phase that destroys completely the correction scheme. 
It is interesting to note that just as in the previous example, a difference between the two decay rates 
does not cause a significant change in the efficiency of the QG.

 \begin{table}
\begin{tabular}{|c|c|c|c|c|}
	\hline
 &  $R^{max}$    &   $R_n^{max}$  & $
\left\langle {R } \right\rangle $   & $
\left\langle {R_n } \right\rangle $
  \\
	\hline
1   & $5.8\times 10^{-5}$ & $5.8\times 10^{-5}$ & $2.5\times 10^{-5}$  & $2.5\times 10^{-5}$ \\
2   & $3.1\times 10^{-6}$ & $3.1\times 10^{-6}$ & $1.9\times 10^{-7}$  & $1.3\times 10^{-7}$ \\
3   & $2.1\times 10^{-1}$ & $1.9\times 10^{-1}$ & $1.0\times 10^{-1}$  & $9.4\times 10^{-2}$ \\
4   & $3.1\times 10^{-6}$ & $3.1\times 10^{-6}$ & $4.1\times 10^{-7}$  & $1.3\times 10^{-7}$ \\
	\hline
	\end{tabular}
\caption{Deviations from a general target byte for four different test cases (see the text for details). The columns display the maximal and the average of the deviations $R$ and $R_n$ for the four cases. }
\label{results2}
\end{table}

The last case that was checked (in the line with yellow diamonds), is a scrambling of the upper byte transformation, 
which makes a Hadamard transformation to the upper byte consecutive to the regular distilling transformation. 
The result of this mutation is a bearable accumulated error, which is moved constantly to the upper byte 
and leaves the lower byte close to the target state, but with lower norm. 

\section{Discussion and Conclusions}

An integral and crucial part of quantum computing and information research is devoted to quantum error correction (QEC) \cite{Steane2,Shor,Steane1}. 
The main question in QEC is the following: Suppose A is sending quantum information to a receiver B. 
An unavoidable influence of a noise may distort the quantum information with probability $p < 1/2$. 
How would B be able to reconstruct the data that was sent to A?
The solution to this problem is usually given by redundancy. 
Before sending his quantum information, A must duplicate his data in several copies from which B 
would be able to withdraw the original data to a very high accuracy.

The task of building a quantum governor is close to QEC, but is different in both motivation and strategy.
A QG main goal is to reduce the influence of noise on the channel between A and B, that is, to reduce $p$. Moreover, the strategy to 
achieve the control uses mainly the system itself and does not create extra information. Considering the fact that 
the scaling of the difficulty of building quantum computers is believed to be exponentially in the size of the system,
it seems that the task of protecting a single byte from decoherence might be more important for quantum computing then the ability to use QEC.

The target entitled here as Quantum Governor can be stated in two versions:
\begin{itemize}
\item{The frail QG - one or more of the system expectation values is constrained, e.g., energy, angular momentum etc.. This constraint is quite similar to the original constraint imposed by James Watt on his classical governor. The similarities between such QG and error corrections are seemingly minor.}
\item{The robust QG - the task of interest is the full conservation of the state of the system, and not only one of its observables. 
This demand is more difficult then the one imposed by the Watt governor but it brings the robust QG closer to the error correction field.}
\end{itemize}
Due to the fact that uncomutative operators cannot be measured simultaneously, it is well understood that feedback control might be 
applicative to several kinds of frail QG, but surely not for any robust QG.

In this paper the QG problem was solved for the robust case for a model two level system. 
In this case the difference between the two versions is not large. 
The robust QG problem is also related to the problems of refocusing and (dynamical) decoupling\cite{Ban,Viola,Viola1,Khodjasteh}.
Both approaches aim to reduce influences of noise  coupled to a quantum system.
However, the strategies of the two schemes are totally different.
The QG problem is a {\it state} oriented problem. It demands the conservation of known states, 
from the influence of noise of an unknown form.  A treatment of the
noise in terms of stochastic quantum equations (see for example\cite{Wiseman,Mancini,Vitali}) is therefore unnecessary. 
The problems of refocusing and decoupling are {\it noise} oriented problems. They try to immune unknown states from a noise with a known form. Accordingly, the solution to the last couple of problems, e.g., Bang-Bang methods and its derivatives,   
uses the known structure of the noise in order to build the appropriate immune decoupling.

To summarize, in this paper the fundamental demands for the task of building a Quantum Governor were developed.
The basis was set for a scheme to achieve automatic control on simple quantum systems. 
The scheme was demonstrated through simulations  on two level system. 
A reduction of the noise by more then two orders of magnitude was achieved. 
The necessary features of a working QG under the present scheme were examined.

The extension of the present  scheme to more complicated system requires additional study. 
Several other schemes and methods to achieve QG, for example the use of the Quantum Zeno paradox could be purposed. 
The exploration of these possibilities is still under investigation. 

\section*{Aknowledgements}
Work supported by the Israel Science foundation. 
The authors would like to thank Jose P. Palao for his assistance, 
and Daniel Lidar for the fruitful discussions.

\end{document}